\documentclass[onecolumn,aps,prd,preprintnumbers,showpacs,superscriptaddress,nofootinbib,amsmath,amssymb,floats,floatfix,showkeys,notitlepage,longbibliography]{revtex4-1}

\usepackage[utf8]{inputenc}
\usepackage{graphicx}
\usepackage{dcolumn}
\usepackage[dvipsnames]{xcolor}
\usepackage[T1]{fontenc}

\usepackage{mathrsfs}  
\usepackage{cases}
\usepackage{bm}
\usepackage{academicons}
\usepackage{mathtools, nccmath}
\usepackage{fancyhdr}
\usepackage{tikz,xcolor}

\usepackage{tensor}
\usepackage{orcidlink}
\usepackage[normalem]{ulem}
\usepackage{lipsum}
\usepackage{soul}
\usepackage{cancel}
\usepackage{stackengine,scalerel}
\usepackage{mathabx}
\usepackage{hyperref}
\usepackage{tabularx}
\hypersetup{colorlinks, linkcolor={red},citecolor={blue},urlcolor={blue}}

\definecolor{lime}{HTML}{A6CE39}
\DeclareRobustCommand{\orcidicon}{
	\begin{tikzpicture}
	\draw[lime, fill=lime] (0,0) 
	circle [radius=0.16] 
	node[white] {{\fontfamily{qag}\selectfont \tiny ID}};
	\draw[white, fill=white] (-0.0625,0.095) 
	circle [radius=0.007];
	\end{tikzpicture}
	\hspace{-2mm}
}

\fancypagestyle{plain}{%
  \fancyhf{}
  \fancyfoot[C]{\iffloatpage{}{\thepage}}
  }
\foreach \x in {A, ..., Z}{%
	\expandafter\xdef\csname orcid\x\endcsname{\noexpand\href{https://orcid.org/\csname orcidauthor\x\endcsname}{\noexpand\orcidicon}}
}

\begin{document}

\title{Exact Dynamical Regular Black Holes from Generalized Polytropic Matter}

\author{Dmitriy Kudryavcev 
\orcidlink{0009-0003-1929-0116}}
\email{kudryavtsiev33@gmail.com}
\affiliation{Department of High Energy and Elementary Particles Physics, Saint Petersburg State University, University Emb. 7/9, Saint Petersburg, 199034, Russia}

\author{Yi Ling
\orcidlink{0000-0001-7392-8641}
}
\email{lingy@ihep.ac.cn}
\affiliation{School of Physical Sciences, University of Chinese Academy of Sciences, Beijing 100049, China}
\affiliation{Institute of High Energy Physics, Chinese Academy of Sciences, Beijing 100049, China}

\author{Vitalii Vertogradov
\orcidlink{0000-0002-5096-7696}}
\email{vdvertogradov@gmail.com}
\affiliation{Physics department, Herzen state Pedagogical University of Russia,
48 Moika Emb., Saint Petersburg 191186, Russia.}
\affiliation{Center for Theoretical Physics, Khazar University, 41 Mehseti Street, Baku, AZ-1096, Azerbaijan.}
\affiliation{SPB branch of SAO RAS, 65 Pulkovskoe Rd, Saint Petersburg
196140, Russia.}

\begin{abstract}
We present a class of exact, dynamical, and fully analytic solutions describing regular black holes formed via the gravitational collapse of matter obeying a generalized polytropic equation of state. Starting from a Vaidya-type geometry with a radially dependent mass function, we demonstrate that regularization of the Kiselev solutions can be achieved through a physically motivated modification of the energy density profile. This procedure leads to nonsingular spacetimes with a de~Sitter core and finite curvature invariants at the center.

We show that the resulting matter content is naturally described by a generalized polytropic equation of state of the form $P=\alpha\rho-\zeta\rho^{\gamma}$, where the polytropic index $\gamma$ is uniquely determined by the regularization scheme. Within this framework, we obtain exact dynamical generalizations of several well-known regular black hole solutions, including the Hayward and Bardeen spacetimes, as particular cases corresponding to specific values of the polytropic parameters.

Remarkably, the requirement that the equation of state remains coordinate independent imposes a universal constraint relating the regularization scale to the mass function, which in turn guarantees the existence of a regular de~Sitter core with a curvature scale independent of the black hole mass. Our results provide a unified analytic description of Hayward-like and Bardeen-like black holes emerging from gravitational collapse, offering a consistent effective-matter interpretation rooted in generalized polytropic matter.
\end{abstract}

\date{\today}



\maketitle
\section{Introduction}

Black holes occupy a central position in modern gravitational physics and astrophysics. Over the past decade, their status has evolved from purely theoretical constructs to firmly established astrophysical objects. The direct detection of gravitational waves by the LIGO and Virgo collaborations, beginning with the landmark observation of GW150914~\cite{bib:abbott2016gw150914}, has provided compelling evidence for the existence of binary black hole systems and has opened an entirely new observational window onto the strong-field regime of gravity. Subsequent detections have revealed a rich population of black holes with a wide range of masses and spins, confirming key predictions of general relativity and offering unprecedented tests of gravitational dynamics in the highly nonlinear regime.

An independent and equally striking confirmation of black hole physics came from the Event Horizon Telescope (EHT) collaboration, which in 2019 produced the first horizon-scale image of the supermassive black hole in the galaxy M87~\cite{bib:eht2019m87}, followed by the image of Sagittarius~A$^\ast$ at the center of our own Galaxy~\cite{bib:eht2022sgrA}. These observations directly probe the near-horizon geometry and the photon capture region, providing empirical access to spacetime curvature on scales of order the Schwarzschild radius. Together, gravitational-wave astronomy and horizon-scale imaging have firmly established black holes as real physical entities and have motivated renewed interest in understanding their internal structure and formation mechanisms.

Despite these successes, classical black hole solutions of general relativity suffer from a profound conceptual difficulty: the presence of spacetime singularities. According to the singularity theorems of Hawking and Penrose~\cite{bib:hawking1970singularities, bib:penrose}, under very general conditions - including reasonable energy conditions and global assumptions - gravitational collapse inevitably leads to the formation of singularities, where curvature invariants diverge and the classical description breaks down. While cosmic censorship conjectures suggest that such singularities are hidden behind event horizons and therefore unobservable to distant observers, their existence signals an incompleteness of classical general relativity and points to the necessity of new physics in regimes of extreme curvature.

Moreover, numerous examples exist in the literature where gravitational collapse leads to the formation of so-called naked singularities~\cite{bib:joshi, bib:joshi_review, bib:vertogradov2016gc, bib:vertogradov2022ijmpa, bib:vertogradov2018ijmpa, bib:dei, bib:inhomo1, bib:inhomo2}. Furthermore, there is research indicating that naked singularities, like black holes, can cast shadows~\cite{bib:joshi_shadow, bib:vertogradov2024epjc}, making it impossible to distinguish a naked singularity from a black hole based on its shadow alone. However, the very presence of a singularity suggests that General Relativity is fundamentally inadequate for describing the spacetime region near such a dense object. Consequently, the scientific community has embarked on a search for new solutions to Einstein's field equations that could circumvent the conditions of Penrose's theorem and yield a black hole without a singularity - the so-called regular black hole.
Historically, several approaches have been proposed to address the singularity problem. One of the earliest and most influential ideas in this direction was put forward by Sakharov and Gliner~\cite{bib:sakharov, bib:gliner}. They proposed that, at sufficiently high (critical) energy densities, ordinary matter may undergo a transition into a vacuum-like state characterized by an effective equation of state $P \simeq -\rho$. In a gravitational context, such a phase naturally gives rise to a de~Sitter–like core, providing a possible mechanism for avoiding curvature singularities during gravitational collapse. This concept laid the conceptual foundation for many subsequent models of regular black holes, in which a de~Sitter interior replaces the classical singularity and smoothly connects to an asymptotically Schwarzschild or Schwarzschild-like exterior. One line of thought invokes quantum gravity effects, which are expected to become important at Planckian densities and may resolve singularities through fundamentally new degrees of freedom or discreteness of spacetime~\cite{bib:ashtekar2005quantum, bib:rovelli2004quantum}. Another approach, more phenomenological in nature, seeks to construct effective classical spacetimes that are nonsingular by modifying either the gravitational dynamics or the matter content at high densities. Within this latter perspective, the idea of \emph{regular black holes} has played a prominent role.

Regular black holes are solutions of Einstein’s equations (possibly coupled to modified matter sources) that possess an event horizon but are free of curvature singularities everywhere. The earliest example was proposed by Bardeen~\cite{bib:bardeen1968non}, who introduced a static, spherically symmetric black hole with a de~Sitter-like core. Although the original Bardeen solution was not derived from a specific matter Lagrangian, it was later shown that it can be interpreted as arising from nonlinear electrodynamics coupled to gravity~\cite{bib:ayonbeato2000bardeen}. Subsequently, Hayward~\cite{bib:hayward2006formation} proposed a particularly simple model of a regular black hole with a de~Sitter core smoothly matched to an asymptotically Schwarzschild exterior.  Later, a sort of regular black holes with Minkowskian core was also constructed~\cite{Xiang:2013sza,Simpson:2019mud,Ling:2021olm}.  Many further regular solutions have since been constructed, often relying on nonlinear electrodynamics, effective anisotropic fluids, or phenomenological energy-momentum tensors~\cite{bib:bronnikov2001regular, bib:dymnikova1992vacuum, bib:ali2024cqg, bib:rincon2024podu, bib:khlopov, bib:khlopov2, bib:khlopov3, bib:khlopov4, bib:khlopov5, bib:baptista2024}.

While these models successfully eliminate curvature singularities, they raise a fundamental physical question: \emph{how can such regular black holes form in realistic gravitational collapse scenarios?} Astrophysical black holes are believed to originate from the collapse of baryonic matter-ordinary matter composed of fermions and bosons obeying well-tested equations of state. In contrast, many regular black hole models rely on exotic matter sources with properties that are difficult to reconcile with known microphysics, such as violations of standard energy conditions or matter content that cannot plausibly exist inside collapsing stars. This tension between mathematical regularity and physical plausibility represents a major obstacle in interpreting regular black holes as realistic astrophysical objects.

The problem of dynamical formation of regular black holes has therefore attracted significant attention. One possible resolution is that effective equations of state describing baryonic matter may change qualitatively at ultrahigh densities, for instance due to strong interactions, phase transitions, or collective effects. In such regimes, the matter may develop significant pressure contributions that counteract further collapse and generate an effective de~Sitter-like core. This idea is conceptually appealing, as it does not require fundamentally exotic matter, but rather an effective description of known matter under extreme conditions. However, constructing explicit, fully analytic, and dynamical solutions realizing this scenario remains a challenging task.
A mechanism for the formation of regular black holes during gravitational collapse has recently been proposed. This mechanism involves baryonic or quark matter undergoing phase transitions at high densities, transforming into a new form of matter. As a result, a significant amount of energy is released in the form of electromagnetic radiation~\cite{bib:vertogradov2025podu, bib:ali2025jcap, vertogradov2025epjc, bib:daniil2025cpc}. This radiation could potentially be detected by an external observer. Crucially, the specific type of matter formed via this process would determine the precise amount of energy released. Therefore, in principle, we could infer the internal composition of the resulting black hole by observing the dynamics of its formation.

In this work, we address precisely this issue. We study the gravitational collapse of baryonic matter described by a generalized polytropic equation of state and demonstrate that it naturally leads to a broad class of exact, dynamical, and regular black hole solutions. Our approach is based on Vaidya-type geometries with a mass function depending on an advanced time coordinate and the radial coordinate. Starting from the Kiselev class of solutions~\cite{bib:husain1996exact, bib:kiselev2003quintessence, bib:tur1, bib:tur2, bib:tur3}, which describe black holes surrounded by barotropic matter, we introduce a physically motivated regularization of the energy density profile. This regularization ensures that the mass function vanishes at the center, thereby eliminating curvature singularities and producing a de~Sitter core.

A key result of our analysis is that the regularized solutions are supported by an effective equation of state of generalized polytropic form,
\[
P = \alpha \rho - \zeta \rho^\gamma,
\]
where the nonlinearity encoded in the second term becomes dominant at high densities and is responsible for the regularization of the geometry. Remarkably, the requirement that the equation of state be coordinate independent imposes a universal constraint relating the regularization scale to the mass function. This condition guarantees not only regularity but also the existence of a universal central curvature scale, independent of the black hole mass.

Within this unified framework, several well-known regular black hole spacetimes arise as special cases. In particular, we show that the Hayward metric emerges uniquely in the case of gravitational collapse of a stiff fluid, corresponding to a specific value of the polytropic parameter. For other types of baryonic matter, the resulting spacetimes are more intricate and lead to what we refer to as \emph{generalized Hayward solutions}. Similarly, Bardeen-like geometries arise for different regularization profiles of the energy density, again governed by the same underlying polytropic structure. In this sense, Hayward and Bardeen black holes are not isolated constructions, but rather particular realizations within a broader family of exact dynamical solutions sourced by generalized polytropic matter.

The purpose of this article is therefore twofold. First, we provide a systematic and physically motivated method for constructing regular black holes from gravitational collapse of baryonic matter, without invoking ad hoc exotic sources. Second, we offer a unified analytic description of Hayward-like and Bardeen-like spacetimes, clarifying their relation to generalized equations of state and highlighting the special role played by stiff matter in the emergence of the original Hayward solution.

The paper is organized as follows. In Section~2, we discuss the general properties of an arbitrary spherically symmetric regular black hole and formulate the basic regularity conditions imposed on the spacetime geometry and matter content. Section~3 is devoted to a transparent presentation of the regularization method based on the Kiselev metric; in this section, we also apply the method to the charged Vaidya spacetime and demonstrate how regularization can be implemented in a dynamical setting.

In Section~4, we derive a new class of regular black hole solutions that reduce to the Kiselev solution in the appropriate limit and can be interpreted as a generalized version of the Hayward spacetime. We further show that the original Hayward solution arises as a particular case within this broader class, with its emergence being dictated by the specific equation of state of the collapsing matter. Section~5 applies the same regularization strategy to the gravitational collapse of baryonic matter leading to the Bardeen metric. In this case, we also obtain a wider family of solutions which, in contrast to the generalized Hayward case, exhibits a number of conceptual and physical drawbacks.

In Section~6, we extract general features common to all three classes of models considered and demonstrate that all regular solutions obtained in this work originate from the same underlying mechanism: at sufficiently high densities, the equation of state of baryonic matter effectively transforms into a generalized polytropic form. Section~7 is dedicated to elucidating the physical nature of the regularization parameter $\beta$. In Section~8, we present an explicit model of gravitational collapse that dynamically leads to the formation of the regular black holes constructed in this paper. Finally, Section~10 contains our conclusions and a discussion of the results.

Throughout this work, we use geometrized units $G = c = 1$ and adopt the spacetime signature $(-,+,+,+)$.

\section{Regular Black Hole Description}

In this section, we briefly describe the conditions that a line element must satisfy in order to represent a spherically symmetric regular black hole.

We consider a spherically symmetric dynamical metric describing a black hole of the form
\begin{equation} \label{eq:metric}
ds^2 = -f(v,r)\,dv^2 + 2\,dv\,dr + r^2 d\Omega^2,
\end{equation}
where the lapse function is given by
\begin{equation}
f(v,r) = 1 - \frac{2M(v,r)}{r},
\end{equation}
and $M(v,r)$ is the mass function depending on the advanced time $v$ and the radial coordinate $r$. Here,
\[
d\Omega^2 = d\theta^2 + \sin^2\theta\, d\varphi^2
\]
denotes the metric on the unit two-sphere, while $v$ is the Eddington--Finkelstein advanced time.

Applying the Einstein equations to the metric \eqref{eq:metric} with an energy--momentum tensor of an anisotropic fluid,
\begin{equation}
T_{ik} = T_{ik}^{(m)} + T_{ik}^{(r)},
\end{equation}
we assume that $T_{ik}^{(m)}$ corresponds to the matter distribution and has the form
\begin{equation} \label{eq:emtgeneralized}
T^{(m)}_{ik} = (\rho + P)(l_i n_k + l_k n_i) + P g_{ik},
\end{equation}
where $\rho$ and $P$ are the energy density and pressure of the matter content, respectively. The vectors $l^i$ and $n^i$ are two null vectors satisfying
\begin{equation}
n^i n_i = 0, \qquad n^i l_i = -1.
\end{equation}
In the coordinates associated with the metric \eqref{eq:metric}, these null vectors take the explicit form
\begin{eqnarray}
l_i &=& \delta^0_i, \nonumber \\
n_i &=& \frac{1}{2}\left(1 - \frac{2M}{r}\right)\delta^0_i - \delta^1_i.
\end{eqnarray}

The component $T^{(r)}_{ik}$ describes the energy flux and is given by
\begin{equation} \label{eq:emtvaidya}
T^{(r)}_{ik} = \sigma(v,r)\, l_i l_k,
\end{equation}
where $\sigma(v,r)$ is the radiation energy flux density.

The Einstein equations with the matter content given by the combination of \eqref{eq:emtgeneralized} and \eqref{eq:emtvaidya} lead to the following relations:
\begin{eqnarray} \label{eq:dengen}
\sigma &=& 2\frac{\dot{M}}{r^2}, \nonumber \\
\rho &=& \frac{2M'}{r^2}, \nonumber \\
P &=& -\frac{M''}{r},
\end{eqnarray}
where a dot and a prime denote derivatives with respect to $v$ and $r$, respectively.

To ensure that the black hole is regular throughout spacetime, we evaluate the curvature invariants: the Ricci scalar $R$, the square of the Ricci tensor $S = R_{ik}R^{ik}$, and the Kretschmann scalar $K = R_{iklm}R^{iklm}$. In terms of the mass function $M(v,r)$, these invariants are given by
\begin{eqnarray} \label{eq:curvature}
R &=& \frac{4M' + 2rM''}{r^2}, \nonumber \\
S &=& \frac{8M'^2 + 2r^2 M''^2}{r^4}, \nonumber \\
K &=& \frac{48M^2 - 64rMM' + 32r^2 M'^2 + 16r^2 MM'' - 16r^3 M' M'' + 4r^4 M''^2}{r^6}.
\end{eqnarray}

Alternatively, these curvature invariants can be expressed in terms of the mass function $M(v,r)$, the energy density $\rho$, and the pressure $P$ from \eqref{eq:dengen} as
\begin{eqnarray} \label{eq:curvature2}
R &=& 2\rho - 2P, \nonumber \\
S &=& 2\rho^2 + 2P^2, \nonumber \\
K &=& \frac{48M^2}{r^6} - \frac{16M}{r^3}(2\rho - P) + 8\rho^2 - 8\rho P + 4P^2.
\end{eqnarray}

Thus, the absence of curvature singularities inside the black hole spacetime can be formulated in terms of the following three conditions:
\begin{enumerate}
\item The energy density must remain finite everywhere. In particular,
\begin{equation}
\lim_{r \to 0} \rho = \rho_0,
\end{equation}
where $\rho_0$ is a finite constant.
\item The matter pressure must also be finite throughout spacetime,
\begin{equation}
\lim_{r \to 0} P = P_0,
\end{equation}
with $P_0$ finite.
\item The mass function must vanish at the center,
\begin{equation}
\lim_{r \to 0} M(v,r) = 0.
\end{equation}
\end{enumerate}

Several remarks are in order. First, a regular black hole is said to possess a de Sitter core if the following condition holds:
\begin{equation}
\lim_{r \to 0} P = - \lim_{r \to 0} \rho.
\end{equation}
Second, it is not necessary to impose additional constraints on the first and second derivatives of the mass function, since their regularity is already guaranteed by the finiteness of the energy density and pressure. Therefore, it is sufficient to require that the mass function vanishes at the center.

\section{Regularization of the Kiselev Solution}

As in the previous section, we assume that the line element describing a spherically symmetric black hole is given by
\begin{equation} \label{eq:metric1}
ds^2 = -\left(1-\frac{2M(v,r)}{r}\right)dv^2 + 2\varepsilon\, dv\, dr + r^2 d\Omega^2,
\end{equation}
where $M(v,r)$ is the mass function depending on both the radial coordinate $r$ and the Eddington time $v$, $\varepsilon = \pm 1$ corresponds to ingoing or outgoing energy flux, and $d\Omega^2 = d\theta^2 + \sin^2\theta\, d\varphi^2$ is the metric on the unit two-sphere. Without loss of generality, we assume $\varepsilon = +1$, since we are primarily interested in black hole formation.

The spacetime described by the metric~\eqref{eq:metric1} has been extensively explored in the literature in the context of gravitational collapse and dynamical black hole formation; see, for instance, Refs.~\cite{bib:Maombi1, bib:Maombi2, bib:maxim2022universe, bib:r2}. Various geometric and physical aspects of this class of metrics, including conformal symmetries, spacetime embeddings, and related properties, have been analyzed in Refs.~\cite{bib:maharaj_conformal, bib:charged_conformal, bib:vertogradov2023mpla, bib:r4, bib:r1, bib:r3}. Regular and nonsingular black hole configurations within closely related frameworks have also been discussed in a number of works, see e.g.~\cite{bib:non_singular}.

The generalized Vaidya geometry is characterized by the presence of an off-diagonal term in the metric, which, in principle, may give rise to negative-energy states for test particles, in close analogy with the situation encountered in the Kerr spacetime~\cite{bib:pen}. It has been shown, however, that such states are absent for neutral particles in the generalized Vaidya background~\cite{bib:vertogradov2020universe}. Nevertheless, an analogous phenomenon can occur for charged particles, allowing for a generalized Penrose-type energy extraction process. Such mechanisms have been investigated both in the static Reissner--Nordström spacetime~\cite{bib:rufini} and in the dynamical Bonnor-Vaidya case~\cite{bib:ver_penrose}.

The physical quantities associated with the spacetime \eqref{eq:metric1} are given by
\begin{eqnarray} \label{eq:density}
\sigma(v,r) &=& \frac{2\dot{M}(v,r)}{r^2}, \nonumber \\
\rho(v,r) &=& \frac{2M'(v,r)}{r^2}, \nonumber \\
P(v,r) &=& -\frac{M''}{r}.
\end{eqnarray}
Here $\sigma$ is associated with the energy flux density, while $\rho$ and $P$ denote the energy density and pressure of the matter, respectively.

Assuming that the energy density and pressure are related by a barotropic equation of state,
\begin{equation}
P = \alpha \rho,
\end{equation}
where $\alpha \in [-1,1]$ and $\alpha \neq \frac{1}{2}$, one arrives at the Husain solution~\cite{bib:husain1996exact}
\begin{equation} \label{eq:husain}
M(v,r) = M_0(v) + D(v)\, r^{1-2\alpha}.
\end{equation}
Here $M_0(v)$ is associated with the black hole mass, while $D(v)$ may be related to a cosmological constant (for $\alpha = -1$) or to a combination of electric and magnetic black hole charges (for $\frac{1}{2} < \alpha \leq 1$)~\cite{Vertogradov:2025gqb}.

It is important to note that the Husain metric \eqref{eq:husain} effectively describes barotropic-like matter. This follows from the fact that the structure of the original metric implies linearity and additivity of the Einstein tensor and, consequently, of the energy--momentum tensor. This property is reflected in the conditions
\[
G^0_{\;0} = G^1_{\;1}, \qquad G^2_{\;2} = G^3_{\;3},
\]
and
\[
T^0_{\;0} = T^1_{\;1}, \qquad T^2_{\;2} = T^3_{\;3}.
\]

This, in turn, leads to anisotropic pressure, since in general $T^1_{\;1} = P_r \neq T^2_{\;2} = P_t$. The only case in which the pressure becomes isotropic corresponds to the equation of state $P = -\rho$, which yields the Kottler solution (also known as the Schwarzschild--de~Sitter solution).

An attempt to impose an isotropic character on the matter distribution was made by Kiselev. He postulated that the barotropic equation of state should take the form $\bar{P} = \omega \rho$, where the averaged pressure is defined as
\[
\bar{P} = \frac{1}{3}\left(P_r + 2P_t\right).
\]
This assumption leads to a relation between the equation-of-state parameters $\alpha$ and $\omega$,
\begin{equation} \label{eq:kis_hus}
\alpha = \frac{1}{2}\left(3\omega + 1\right).
\end{equation}

Thus, for matter satisfying the dominant energy condition with positive tangential pressure $P_t > 0$, the allowed values of the parameter $\omega$ lie in the interval $(0,1]$. This corresponds to the condition $\alpha \in \left(\frac{1}{2}, 2\right]$ for the parameter $\alpha$.

Since the functional forms of the Kiselev and Husain metrics are identical and differ only through the constraint \eqref{eq:kis_hus}, we shall use the parameter $\alpha$ in what follows, keeping in mind that its admissible values are restricted to the half-open interval $\alpha \in \left(\frac{1}{2}, 2\right]$.

The energy density and pressure corresponding to the solution \eqref{eq:husain} take the form
\begin{eqnarray}
\rho &=& 2\left(1-2\alpha\right)\frac{D(v)}{r^{2\alpha+2}}, \nonumber \\
P &=& 2\alpha\left(1-2\alpha\right)\frac{D(v)}{r^{2\alpha+2}}.
\end{eqnarray}
The weak energy condition requires the positivity of the energy density $\rho$. Therefore, we conclude that $(1-2\alpha)D(v) \geq 0$, which leads to the following constraints:
\begin{eqnarray}
\alpha < \frac{1}{2} &\implies& D(v) > 0, \nonumber \\
\alpha > \frac{1}{2} &\implies& D(v) < 0.
\end{eqnarray}

The solution \eqref{eq:husain} can describe only a singular black hole, since it never satisfies the condition
\[
\lim_{r \to 0} M(v,r) = 0,
\]
which is a key requirement for the existence of a regular center.

In order to regularize the Husain solution \eqref{eq:husain}, one may either employ the methods described in~\cite{bib:frolov,bib:rn_regular} or adopt the approach outlined below. This method consists in introducing a new regularization function $\beta \equiv \beta(v)$ into the energy density,
\begin{equation} \label{eq:density_reg}
\rho(v,r) \rightarrow \rho(v,r+\beta) =
\frac{2(1-2\alpha)D(v)}{(r+\beta)^{2\alpha+2}}.
\end{equation}

Subsequently, solving the differential equation \eqref{eq:density}, one obtains the mass function from
\begin{equation} \label{eq:diferential}
\frac{2M'}{r^2} =
\frac{2(1-2\alpha)D(v)}{(r+\beta)^{2\alpha+2}},
\end{equation}
and imposes an additional condition on the function $D(v)$ such that
\[
\lim_{r \to 0} M(v,r) = 0.
\]

In order to solve Eq.~\eqref{eq:diferential}, three cases must be considered separately:
\begin{enumerate}
\item $\alpha = 0$,
\item $\alpha = -\frac{1}{2}$,
\item the general case excluding the two special cases above, i.e.\ $\alpha \in [-1,1]$ with $\alpha \neq -\frac{1}{2}$, $\alpha \neq 0$, and $\alpha \neq \frac{1}{2}$.
\end{enumerate}
In the following subsections, all these possibilities will be analyzed in detail.

\subsection{The case $\alpha = 0$}

The first particular case that we consider corresponds to $\alpha = 0$. Solving Eq.~\eqref{eq:diferential} for this value of the parameter, one obtains
\begin{equation} \label{eq:dust_regular}
\begin{split}
M(v,r) = M_0(v) + D(v)\bigl(r+\beta(v)\bigr)
- 2\beta(v)D(v)\ln\lvert r+\beta(v)\rvert
- \frac{\beta^2(v)D(v)}{r+\beta(v)} .
\end{split}
\end{equation}

Note that in the limit $\beta(v) \equiv 0$ this solution reduces to the Husain solution with $\alpha = 0$,
\begin{equation}
M(v,r) = M_0(v) + D(v) r, \qquad \beta(v) \equiv 0.
\end{equation}

The energy density and pressure corresponding to the solution \eqref{eq:dust_regular} are given by
\begin{eqnarray}
\rho &=& \frac{2D(v)}{(r+\beta(v))^2}, \nonumber \\
P &=& -\frac{2D(v)\beta(v)}{(r+\beta(v))^3}.
\end{eqnarray}
Note that for $D(v) > 0$ the weak energy condition is satisfied everywhere in spacetime. The equation of state reads
\begin{equation}
P = -\frac{\beta}{r+\beta}\,\rho .
\end{equation}
In the limit $\beta \rightarrow 0$, one finds $P \rightarrow 0$, i.e.\ the original $\alpha = 0$ case is recovered. Moreover, since $\frac{\beta}{r+\beta} \leq 1$, the dominant energy condition is also satisfied throughout spacetime. The strong energy condition, however, is violated near the center, which is a generic feature of regular black hole models.

Thus, both the energy density and the pressure remain finite at the center, and the only remaining condition required to obtain a non-singular black hole is
\[
\lim_{r \rightarrow 0} M(v,r) = 0 .
\]
Explicitly,
\begin{equation}
\begin{split}
\lim_{r \rightarrow 0} M(v,r)
= M_0(v) - 2\beta(v)D(v)\ln\lvert \beta(v)\rvert = 0
\;\;\Rightarrow\;\;
D(v) = \frac{M_0(v)}{2\beta(v)\ln\lvert\beta(v)\rvert}.
\end{split}
\end{equation}

Since the parameter $\beta$ was introduced as a small regularization parameter, $\beta \ll 1$, this condition implies that $D(v)$ must be negative, which in turn leads to a violation of the weak energy condition. For this reason, we regard this model as unphysical and discard it.

\subsection{The case $\alpha = -\frac{1}{2}$}

In this case, Eq.~\eqref{eq:diferential} yields the solution
\begin{equation} \label{eq:regular_negative}
\begin{split}
M(v,r) = M_0(v) + D(v)\bigl(r+\beta(v)\bigr)^2
- 4\beta(v)D(v)\bigl(r+\beta(v)\bigr)
+ \beta^2(v)D(v)\ln\lvert r+\beta(v)\rvert .
\end{split}
\end{equation}

Again, in the limit $\beta \rightarrow 0$ one recovers the Husain solution with $\alpha = -\frac{1}{2}$,
\begin{equation}
M(v,r) = M_0(v) + D(v) r^2 .
\end{equation}

The energy density and pressure corresponding to \eqref{eq:regular_negative} are given by
\begin{eqnarray}
\rho &=& \frac{4D(v)}{r+\beta(v)}, \nonumber \\
P &=& -\frac{2D(v)\bigl(r+2\beta(v)\bigr)}{(r+\beta(v))^2}.
\end{eqnarray}
One can see that in the limit $r \rightarrow 0$ both the energy density and the pressure remain finite. The corresponding equation of state takes the form
\begin{equation}
P = -\frac{1}{2}\frac{r+2\beta}{r+\beta}\,\rho .
\end{equation}
In the limit $\beta \rightarrow 0$, the equation of state $P = -\frac{1}{2}\rho$ is recovered. Note that $D(v) > 0$ and $\frac{r+2\beta}{r+\beta} \leq 2$, which implies that both the weak and the dominant energy conditions are satisfied throughout spacetime.

The last condition that must be imposed in order to obtain a non-singular black hole is
\begin{equation}
\begin{split}
\lim_{r \rightarrow 0} M(v,r)
= M_0(v) - 3D(v)\beta^2(v)
+ D(v)\beta^2(v)\ln\lvert\beta(v)\rvert = 0
\;\;\Rightarrow\;\;
D(v) = \frac{M_0(v)}{\beta^2(v)\bigl(3-\ln\lvert\beta(v)\rvert\bigr)} .
\end{split}
\end{equation}
In this case, the condition $\beta(v) \ll 1$ leads to a positive value of $D(v)$, and the regularization procedure is mathematically consistent.

However, we still regard this solution as physically unrealistic for the following reasons. If one considers the gravitational collapse of a star, the matter content should be described by a realistic equation of state, and no matter obeying the equation of state $P = -\frac{1}{2}\rho$ should be present in the stellar interior at the initial moment of collapse. Such exotic matter may arise dynamically during gravitational collapse as a result of phase transitions~\cite{vertogradov2025epjc}, but not at the initial stage, which is precisely the regime considered in the present work.

\subsection{General case}

In the general case, when $\alpha \in [-1,2]$ with $\alpha \neq 0$ and $\alpha \neq \pm \frac{1}{2}$, solving Eq.~\eqref{eq:diferential} yields
\begin{equation} \label{eq:regular_gen}
\begin{split}
M(v,r) = M_0(v) + D(v)\,[r+\beta(v)]^{1-2\alpha}
+ \frac{(1-2\alpha)\beta(v)D(v)}{\alpha\,[r+\beta(v)]^{2\alpha}}
- \frac{\beta^2(v)D(v)(1-2\alpha)}{(1+2\alpha)\,[r+\beta(v)]^{1+2\alpha}} .
\end{split}
\end{equation}

In the limit $\beta \rightarrow 0$, the solution \eqref{eq:regular_gen} reduces to the Husain solution,
\begin{equation}
M(v,r) = M_0(v) + D(v) r^{1-2\alpha}, \qquad \beta(v) \equiv 0 .
\end{equation}

The energy density and pressure corresponding to the solution \eqref{eq:regular_gen} are given by
\begin{eqnarray} \label{eq:density_gen}
\rho &=& \frac{2(1-2\alpha)D(v)}{[r+\beta(v)]^{2+2\alpha}}, \nonumber \\
P &=& -\frac{(1-2\alpha)D(v)}{[r+\beta(v)]^{2+2\alpha}}
+ \frac{(\alpha+1)(1-2\alpha)D(v)\, r}{[r+\beta(v)]^{3+2\alpha}} .
\end{eqnarray}

The corresponding equation of state can be written as
\begin{equation} \label{eq:state}
P = \frac{\alpha r - \beta}{r+\beta}\,\rho .
\end{equation}
In the limit $\beta \rightarrow 0$, Eq.~\eqref{eq:state} reduces to the standard barotropic equation of state $P = \alpha \rho$.

However, Eq.~\eqref{eq:state} is written under the implicit assumption that $P/\rho = w(r)$, which is problematic since the equation-of-state parameter depends explicitly on the radial coordinate. Such a dependence generally violates the requirement of general covariance at the level of the matter Lagrangian.

Therefore, it is necessary to rewrite the equation of state in an invariant form. Noting that
\begin{equation}
r = \left[ \frac{2(1-2\alpha)D(v)}{\rho} \right]^{\frac{1}{2+2\alpha}} - \beta ,
\end{equation}
we obtain an equation of state of the form
\begin{equation} \label{eq:state2}
P = \alpha \rho - \zeta \rho^{1+\frac{1}{2\alpha+2}}, \qquad \zeta > 0 ,
\end{equation}
where
\begin{equation} \label{eq:defzeta}
\zeta \equiv \frac{(\alpha+1)\beta(v)}{\left[2(1-2\alpha)D(v)\right]^{\frac{1}{2\alpha+2}}} .
\end{equation}

We note that the equation of state \eqref{eq:state2} represents a generalized polytropic equation of state,
\begin{equation}
P = \alpha \rho - \zeta \rho^{\gamma},
\end{equation}
with
\begin{equation}
\gamma = 1 + \frac{1}{N},
\end{equation}
where $N$ is the polytropic index. The equation-of-state parameter $\omega$ (or $\alpha$) is related to the polytropic index by $\omega = \frac{N-3}{3}$ and $\alpha = \frac{N-2}{2}$. Some properties of polytropic equation of state in connection to gravitational collapse of generalized Vaidya spacetime have been considered in paper~\cite{bib:vertogradov2024grg}.

We must also emphasize that the functions $\beta(v)$ and $D(v)$ must satisfy the condition
\begin{equation}
\frac{\beta(v)}{D^{\frac{1}{2\alpha+2}}(v)} \equiv \text{const}.
\end{equation}
This requirement follows from the same fundamental considerations that forbid the equation-of-state parameters from having an explicit time dependence.

Recalling that $\alpha \in [-1,2]$, one readily verifies that $\rho \geq |P|$, i.e.\ the dominant energy condition is satisfied everywhere in spacetime. We further require that the weak energy condition holds globally. For this purpose, we distinguish two cases:
\begin{itemize}
\item $\alpha < \frac{1}{2}$. In this case, the weak energy condition requires $D(v) \geq 0$;
\item $\alpha > \frac{1}{2}$. In this case, the weak energy condition implies $D(v) \leq 0$.
\end{itemize}

As follows from Eq.~\eqref{eq:density_gen}, both the energy density and the pressure remain finite in the limit $r \rightarrow 0$. Hence, in order to obtain a regular black hole, the final condition
\[
\lim_{r \rightarrow 0} M(v,r) = 0
\]
must be satisfied. This yields
\begin{equation} \label{eq:reccon}
D(v) = -\frac{M_0(v)\left(\alpha + 2\alpha^2\right)}{\beta(v)^{1-2\alpha}} .
\end{equation}

Thus, the present regularization scheme is consistent only for the following range of the parameter $\alpha$:
\begin{equation}
\begin{split}
\text{regular black hole} + \text{weak energy condition}
\;\;\Longrightarrow\;\;
\alpha \in \left[-\frac{1}{2}, 0\right) \cup \left(\frac{1}{2}, 2\right] .
\end{split}
\end{equation}

\subsection{Example: charged Vaidya black hole}

As an illustrative example, let us consider the case $\alpha = 1$ (corresponding to $\omega = \frac{1}{3}$). In this case, the Kiselev solution reduces to the Bonnor--Vaidya solution~\cite{bib:bonor}, with
\[
D(v) = -\frac{Q^2(v)}{2}.
\]
The corresponding metric takes the form
\begin{equation}
ds^2 = -\left(1 - \frac{2M_0(v)}{r} + \frac{Q^2(v)}{r^2}\right) dv^2 + 2\,dv\,dr + r^2 d\Omega^2 .
\end{equation}

Applying the regularization procedure described above, we obtain the generalized line element
\begin{eqnarray}
ds^2 &=& -f(v,r)\,dv^2 + 2\,dv\,dr + r^2 d\Omega^2 ,
\end{eqnarray}
where
\begin{eqnarray}
f(v,r) &=& 1 - \frac{2M_0(v)}{r} \nonumber \\
&+& \frac{3Q^2(v) r^2 + 3Q^2(v)\beta(v) r + Q^2(v)\beta^2(v)}{3\,[r+\beta(v)]^3} .
\end{eqnarray}

For $\beta(v) \equiv 0$, this metric coincides with the charged Bonnor--Vaidya black hole. However, the spacetime describes a regular black hole if and only if the function $\beta(v)$ satisfies the condition
\begin{equation}
\beta(v) = \frac{Q^2(v)}{6M_0(v)} .
\end{equation}

Substituting this expression back into the metric function, we finally arrive at the regular Bonnor--Vaidya spacetime,
\begin{equation}
f(v,r) = 1 - \frac{2M_0(v)}{r}
+ \frac{3Q^2(v) r^2 + \dfrac{Q^4(v) r}{2M_0(v)} + \dfrac{Q^6(v)}{36M_0^2(v)}}
{3r\left[r + \dfrac{Q^2(v)}{6M_0(v)}\right]^3} .
\end{equation}

This example demonstrates explicitly how the proposed construction allows one to obtain a time-dependent, charged black hole geometry that is free of curvature singularities at the origin, while smoothly reducing to the standard Bonnor--Vaidya solution in the appropriate limit. It therefore provides a concrete realization of a regular, dynamical black hole supported by physically motivated matter sources.
\section{Towards the Hayward Spacetime}

We now demonstrate that a generalized Hayward metric can be obtained through a similar procedure by studying the gravitational collapse of baryonic matter. Consider an energy density of the form
\begin{equation}
    \rho = \frac{2\left(1 - 2\alpha\right) D(v)}{\left(r^3 + \beta^3\right)^{\frac{2\alpha + 2}{3}}},
\end{equation}
where \( \beta \) is a regularization parameter. In the limit \( \beta \to 0 \), this expression reduces to the energy density corresponding to a barotropic equation of state:
\begin{equation}
    \rho = \frac{2\left(1 - 2\alpha\right) D(v)}{r^{2\alpha + 2}}.
\end{equation}

From the Einstein equations, the energy density relates to the mass function \( M(v,r) \) via
\begin{equation}
    \rho = \frac{2 M'}{r^2},
\end{equation}
which yields the mass function
\begin{equation} \label{eq:masshaygen}
    M(v,r) = M_0(v) + \frac{D(v)}{\left(r^3 + \beta^3\right)^{\frac{2\alpha - 1}{3}}}.
\end{equation}

Regularity at the origin requires \( M(v, r) \to 0 \) as \( r \to 0 \). Imposing this condition leads to the relation
\begin{equation}
    D(v) \equiv -M_0(v) \, \beta^{\,2\alpha-1}.
\end{equation}
We must recall that for $\alpha>\frac{1}{2}$ the weak energy condition requires $D(v)<0$, which is indeed realized by the expression above.

Substituting this back, we obtain the line element describing a regular black hole:
\begin{equation} \label{eq:genhay}
    ds^2 = -\left(1 - \frac{2M_0(v)}{r} + \frac{2M_0(v)\,\beta^{\,2\alpha-1}}{r\left(r^3 + \beta^3\right)^{\frac{2\alpha - 1}{3}}}\right) dv^2 + 2\,dv\,dr + r^2 d\Omega^2.
\end{equation}

In the limit \( \beta \to 0 \), this metric reduces to the Kiselev black hole solution. Notably, the metric~\eqref{eq:genhay} constitutes a generalization of the Hayward spacetime, which is recovered as a special case when \( \alpha = 2 \) (equivalently, \( \omega = 1 \)), corresponding to an initial stiff fluid configuration. 
It is worth emphasizing that the spacetime~\eqref{eq:genhay} remains regular at the center for all admissible values of $\alpha$, while smoothly interpolating between a de~Sitter--like core and an asymptotically Kiselev geometry, thereby providing a consistent description of regular black hole formation.

Note that the energy density and pressure for the solution \eqref{eq:masshaygen} take the form
\begin{eqnarray}
\rho &=& \frac{2(1 - 2\alpha) D(v)}{\left(r^3 + \beta^3\right)^{\frac{2\alpha + 2}{3}}}, \nonumber \\
P &=& -\frac{2(1 - 2\alpha) D(v)}{(r^3 + \beta^3)^{\frac{2\alpha + 2}{3}}}
      - \frac{(1 - 2\alpha)(2 + 2\alpha) D(v) r^3}{(r^3 + \beta^3)^{\frac{2\alpha + 5}{3}}}.
\end{eqnarray}
with the equation of state
\begin{equation}
P = -\rho + \frac{(\alpha + 1) r^3}{r^3 + \beta^3} \rho.
\end{equation}

Again, the requirement that the coefficients of the equation of state do not depend on the coordinates leads us to a generalized polytropic equation of state:
\begin{equation}
P = \alpha \rho - \zeta \rho^\gamma,
\end{equation}
where
\begin{eqnarray}
\zeta &\equiv & \frac{(\alpha + 1) \beta^3}{\left(2(1 - 2\alpha) D(v)\right)^{\frac{3}{2\alpha + 2}}}, \nonumber \\
\gamma &\equiv & 1 + \frac{3}{2\alpha + 2} = 1 + \frac{1}{\omega}.
\end{eqnarray}

Thus, the polytropic index $N$ is related to the equation-of-state parameter by $\omega = N$ ($\alpha = \frac{3N - 2}{2}$).

Consequently, the functions $\beta(v)$ and $D(v)$ must satisfy the condition
\begin{equation}
\frac{\beta(v)}{D^{\frac{1}{2\alpha + 2}}(v)} = \text{const}.
\end{equation}
\section{Towards the Bardeen spacetime}

Let us now consider a scenario in which the gravitational collapse of baryonic matter leads to the Bardeen solution and to a broader class of related regular geometries.

To this end, we assume the energy density to be of the form
\begin{equation}
\rho = \frac{2(1 - 2\alpha)\, D(v)}{(r^2 + \beta^2)^{\alpha + 1}},
\end{equation}
where, as before, $\beta$ is a regularization parameter controlling the behavior of the matter distribution near the origin. In the limit $\beta \to 0$, this expression reduces to the energy density corresponding to ordinary baryonic matter obeying a barotropic equation of state.

Using the Einstein equation
\[
\rho = \frac{2M'}{r^2},
\]
we obtain the mass function in the integral form
\begin{equation} \label{eq:massbargen}
M(v,r) = M_0(v) + \int \frac{(1 - 2\alpha)\, D(v)\, r^2}{(r^2 + \beta^2)^{\alpha + 1}} \, dr .
\end{equation}
In general, the integral in \eqref{eq:massbargen} can be expressed in terms of hypergeometric functions. However, for particular values of the parameter $\alpha$, namely $\alpha = \frac{3}{2},\,1,\,2$, the integral admits closed-form expressions in terms of elementary functions. 

In particular, for $\alpha = \frac{3}{2}$ one obtains the well-known Bardeen solution,
\begin{equation}
M(v,r) = M_0(v) + \frac{2}{3}\,\frac{D(v)}{\beta}\,
\frac{r^3}{(r^2 + \beta^2)^{3/2}},
\qquad
\alpha = \frac{3}{2},
\qquad
\omega = \frac{2}{3}.
\end{equation}
The original Bardeen solution is recovered under the additional condition
\begin{equation}
M_0(v) \equiv 0,
\qquad
M_{\text{Bardeen}} \equiv \frac{2}{3}\,\frac{D(v)}{\beta}.
\end{equation}

However, the approach employed in this section has an important limitation. In order to reproduce the Bardeen spacetime, one must impose the condition $M_0(v) \equiv 0$, which is the same restriction encountered in the original construction of the Bardeen solution. Moreover, during the integration one typically introduces rescalings of the form $t = r/\beta$. As a consequence, in this case there is no smooth way to recover the Kiselev black hole by taking the limit $\beta \to 0$, in contrast to the Hayward-type regularizations discussed earlier.

The energy density and pressure corresponding to the solution \eqref{eq:massbargen} read
\begin{eqnarray}
\rho &=& \frac{2(1 - 2\alpha)\, D(v)}{(r^2 + \beta^2)^{\alpha + 1}}, \nonumber \\
P &=& -\frac{2(1 - 2\alpha)\, D(v)}{(r^2 + \beta^2)^{\alpha + 1}}
+ \frac{2(\alpha + 1)(1 - 2\alpha)\, D(v)\, r^2}{(r^2 + \beta^2)^{\alpha + 2}} .
\end{eqnarray}
Accordingly, the equation of state can be written in the form
\begin{equation}
P = -\rho + \frac{(\alpha + 1)\, r^2}{r^2 + \beta^2}\,\rho .
\end{equation}

By expressing the pressure solely as a function of the energy density $\rho$, we once again arrive at a generalized polytropic equation of state,
\begin{equation}
P = \alpha \rho - \zeta \rho^{\gamma},
\end{equation}
with
\begin{eqnarray}
\zeta &\equiv& \frac{(\alpha + 1)\, \beta^2}{\left[2(1 - 2\alpha)\, D(v)\right]^{\frac{1}{\alpha + 1}}}, \nonumber \\
\gamma &\equiv& 1 + \frac{1}{\alpha + 1}.
\end{eqnarray}
Requiring that the coefficients of the equation of state be independent of the advanced time coordinate $v$, we again obtain a constraint identical to that found in the previous constructions,
\begin{equation}
\frac{\beta}{D^{\frac{1}{2\alpha + 2}}(v)} = \text{const}.
\end{equation}
In summary, this construction shows explicitly how Bardeen-type regular black holes can be interpreted as arising from specific regularized matter distributions. At the same time, it highlights an essential qualitative difference with respect to the Hayward case: while both geometries are regular at the center, only the Hayward-type solutions allow for a smooth connection to the Kiselev spacetime via a continuous limiting procedure.
\section{Unified density profile and the physical meaning of the regularity constraint}

In this section we collect and generalize the regular matter profiles introduced earlier, and clarify the physical interpretation of the constraint required for a radius-independent equation of state.

\subsection{Unified generalization of the density profile}

In the present work, three different modifications of the energy density $\rho$ have been introduced, namely in Eqs.~(16), (44), and~(55), corresponding to regular Reissner--Nordstr\"om--type, Hayward-type, and Bardeen-type geometries, respectively. In all three cases, the requirement that the equation of state be independent of the radial coordinate $r$ leads to the constraint
\begin{equation}
\frac{\beta(v)}{D(v)^{\,1/(2\alpha+2)}} = \text{const}.
\end{equation}
Remarkably, this condition turns out to be identical for all three regularization schemes.

These apparently different cases can, in fact, be unified within a single generalized ansatz for the energy density,
\begin{equation}
\rho = \frac{2(1 - 2\alpha)\, D(v)}{\bigl(r^{\,n} + \beta(v)^{\,n}\bigr)^{(2\alpha+2)/n}},
\end{equation}
where the values $n = 1,\,3,\,2$ correspond to the regular Reissner--Nordstr\"om--type, Hayward-type, and Bardeen-type solutions discussed in the main text.

For this general class of profiles, the mass function $M(v,r)$ can be expressed in terms of hypergeometric functions. Although the explicit form of $M(v,r)$ is technically involved, the resulting equation of state retains a remarkably simple structure:
\begin{equation}
P = \frac{\alpha r^{\,n} - \beta(v)^{\,n}}{r^{\,n} + \beta(v)^{\,n}}\,\rho
= \alpha \rho
- \frac{(\alpha + 1)\, \beta(v)^{\,n}}
{\bigl[2(1 - 2\alpha)\, D(v)\bigr]^{n/(2\alpha+2)}}
\,\rho^{\,1 + n/(2\alpha+2)} .
\end{equation}
It follows immediately that the condition ensuring the independence of the equation of state from the radial coordinate $r$ is precisely the universal relation
\begin{equation}
\frac{\beta(v)}{D(v)^{\,1/(2\alpha+2)}} = \text{const},
\end{equation}
which therefore plays a central role in all regular black hole configurations considered here.

\subsection{Physical interpretation of the constraint}

The physical meaning of the condition
\begin{equation}
\frac{\beta(v)}{D(v)^{\,1/(2\alpha+2)}} = \text{const}
\end{equation}
deserves further discussion. The central energy density,
\begin{equation}
\rho_0 \equiv \lim_{r \to 0} \rho \;\propto\; \frac{D(v)}{\beta(v)^{\,2\alpha+2}},
\end{equation}
is not only finite due to this constraint, but also becomes independent of the advanced time coordinate $v$. As a consequence, the Kretschmann scalar $K$ at the center ($r \to 0$) approaches a universal constant that does not depend on the black hole mass $M(v)$.

This feature can be explicitly illustrated for the Hayward-type geometry. For $n = 2$, the constraint implies a linear relation of the form $\beta(v) = C\, M(v)$, and the corresponding Kretschmann scalar of the regular metric at the center ($r = 0$) indeed turns out to be independent of the mass $M$.

In contrast, for $n = 1$, corresponding to the regular Reissner--Nordstr\"om--type geometry, the same requirement leads to the unconventional relation
\begin{equation}
\frac{Q^{3}(v)}{M(v)} = C .
\end{equation}
This implies that, if one demands the maximum value of the Kretschmann scalar to be a universal constant independent of both $M$ and $Q$, then the charge and the mass cannot evolve independently. Instead, their dynamics must be correlated through the above relation, which may have nontrivial implications for the physical interpretation of charged regular black holes and their formation scenarios.

A few remarks should be made concerning the dependence of the regularization parameter $\beta$ on the black hole mass $M_0(v)$ in the case of regular black holes. As can be seen from all the models considered above, the regularization constant $\beta$ is related to the black hole mass $M_0$ by the following relation:
\begin{equation}
\beta = \alpha M_0^{\frac{1}{3}}, \qquad \alpha = \text{const}.
\end{equation}
Only in the case of the Bardeen metric does this condition fail to hold.

Such a dependence follows directly from the requirement that the mass function $M(v, r)$ vanish at the center of the black hole. From the Einstein equations it follows that
\begin{equation}
\frac{2M'}{r^{2}} = \rho \quad \Rightarrow \quad M(v, r) = M_0(v) + \frac{1}{2} \int \rho \, r^{2}  dr .
\end{equation}
Imposing the condition that the mass function $M(v, r)$ goes to zero at the center, we obtain the relation
\begin{equation}
M_0(v) = - \lim_{r \to 0} \frac{1}{2} \int \rho \, r^{2} dr .
\end{equation}
The integral on the right-hand side must tend to some function that depends on the regularization parameter but is independent of the black hole mass $M_0(v)$. Hence, in the most general case, the regularization parameter $\beta$ always turns out to depend on the black hole mass,
\begin{equation}
\beta \equiv \beta\bigl(M_0(v)\bigr).
\end{equation}

The apparent contradiction with the Bardeen metric is resolved as follows. As can be seen from the section where the solution generalizing the Bardeen metric is derived, the requirement $M_0(v) \equiv 0$ arises. This is primarily due to the fact that the Bardeen solution was obtained phenomenologically, rather than by first specifying the matter distribution and then solving the Einstein equations. As our method demonstrates, if the integration constant is not set to zero, a regular solution cannot be obtained because the condition $\lim_{r \to 0} M(v, r) = 0$ cannot be satisfied. Consequently, the regularization parameter $\beta$ is a function of $M_0$ in all cases except when the latter is identically zero.

\section{The Nature of $\beta$}

In this section, we clarify the physical meaning of the parameter $\beta$ and its relation to the near-origin geometry and possible matter sources. For definiteness and clarity, we restrict our analysis to the simplest regular profile,
\begin{equation}
\rho = \frac{2(1-2\alpha)\,D(v)}{(r+\beta)^{2\alpha+2}},
\end{equation}
which corresponds to the case $n=1$ in the unified framework discussed above. This choice allows for a fully transparent analytical treatment. We emphasize, however, that analogous arguments can be carried out for the Bardeen-like ($n=2$) and Hayward-like ($n=3$) profiles as well, although the corresponding expressions become technically more involved and less instructive at the intermediate steps. Importantly, the physical conclusions regarding regularity and the interpretation of $\beta$ remain qualitatively the same in those cases.

\subsection{Relation between $\beta$ and the de Sitter core}

Let us consider the mass function~\eqref{eq:regular_gen} subject to the regularity condition~\eqref{eq:reccon}. Requiring the existence of a de~Sitter core at the center of the black hole implies that the lapse function admits the expansion
\begin{equation} \label{eq:razlogen}
f(r) = 1 - \frac{\Lambda}{3} r^2 + \mathcal{O}\!\left(r^3\right),
\end{equation}
as $r \to 0$. In order to satisfy this condition, we compute the second radial derivative of the lapse function. Using $f(r)=1-2M(r)/r$, one finds
\begin{equation}
f''(r) = \frac{4M'(r)r - 4M(r) - 2M''(r)r^2}{r^3}
= 2\rho + 2P - \frac{4M(r)}{r^3}.
\end{equation}

In the limit $r \to 0$, regularity implies the vacuum-like equation of state $P=-\rho$, and therefore
\begin{equation}
\lim_{r \to 0} f''(r)
= \lim_{r \to 0} \left(-\frac{4M(r)}{r^3}\right).
\end{equation}
Using the explicit form of the mass function and the definition of $D(v)$, we obtain
\begin{equation}
\lim_{r \to 0} \left(-\frac{4M(r)}{r^3}\right)
= \frac{4(1-2\alpha)\,D(v)}{\beta^{2+2\alpha}}
= -\frac{4(1-2\alpha)M_0}{3\beta^3},
\end{equation}
where, in the last equality, the regularity condition~\eqref{eq:reccon} has been used.

Comparing this result with the de~Sitter expansion~\eqref{eq:razlogen}, we arrive at a direct relation between $\beta$ and the effective cosmological constant,
\begin{equation}
\beta = \left(\frac{2M_0}{\Lambda}\right)^{1/3}.
\end{equation}
Thus, $\beta$ acquires a clear geometric meaning: it sets the length scale of the de~Sitter core and controls the maximal curvature attained at the center.

\subsection{Interpretation in nonlinear electrodynamics}

An alternative interpretation of the parameter $\beta$ emerges when the regular solution~\eqref{eq:regular_gen} is supported by nonlinear electrodynamics. To reconstruct the corresponding Lagrangian, we employ the standard inverse-engineering method developed in~\cite{bib:ingenering}, according to which
\begin{equation}
\frac{M'(r)}{r^2} = -\mathcal{L}(\mathcal{F}),
\end{equation}
where the electromagnetic invariant is defined as
\begin{equation}
\mathcal{F} = F^{ik}F_{ik} = 2\left(B^2 - E^2\right).
\end{equation}
The magnetic field generated by a monopole of charge $P$ takes the form
\begin{equation}
B = \frac{P}{r^2}.
\end{equation}
Since a regular center is incompatible with a nonvanishing electric field~\cite{bib:bronnikov2001regular}, we set $E\equiv 0$. This allows us to express the radial coordinate in terms of $\mathcal{F}$ as
\begin{equation} \label{eq:emprom}
r = \left(\frac{2P^2}{\mathcal{F}}\right)^{1/4}.
\end{equation}

Substituting the mass function~\eqref{eq:regular_gen} and using~\eqref{eq:emprom}, we obtain the nonlinear electrodynamics Lagrangian
\begin{equation} \label{eq:lagrangian}
\mathcal{L}(\mathcal{F})
= -\frac{(1-2\alpha)\,D(v)\,\mathcal{F}^{(1+\alpha)/2}}
{\left[(2P^2)^{1/4} + \beta\,\mathcal{F}^{1/4}\right]^{2+2\alpha}}.
\end{equation}

Requiring the Maxwell weak-field limit,
\begin{equation}
\mathcal{L} \simeq -\frac{1}{4}\mathcal{F}
+ \mathcal{O}\!\left(\mathcal{F}^2\right),
\end{equation}
fixes the parameter $\alpha$ to be $\alpha=1$, which in turn yields
\begin{equation}
D(v) \equiv -\frac{P^2}{2}.
\end{equation}
Finally, invoking once more the regularity condition~\eqref{eq:reccon}, we obtain
\begin{equation}
\beta = \frac{P^2}{6M_0}.
\end{equation}

We thus conclude that the parameter $\beta$ admits a dual interpretation: geometrically, it determines the size of the de~Sitter core and the maximal curvature scale, while physically, in the context of nonlinear electrodynamics, it is directly related to the magnetic monopole charge supporting the regular black hole.
\section{The Process of de Sitter Core Formation}

The preceding discussion demonstrates how we can regularize the Husain solution using the parameter $\beta$. Furthermore, we can associate the parameter $\beta$ either with the cosmological constant $\Lambda$ or with the magnetic charge of a monopole $P$. However, these arguments do not explain how the de Sitter core forms during the gravitational collapse of baryon-like matter.

During the gravitational collapse of a massive star, as certain densities are reached, matter transitions into a different type. This process is accompanied by energy release in the form of electromagnetic radiation. As shown in Refs.~\cite{bib:vertogradov2025podu, bib:ali2025jcap}, the gravitational collapse of baryonic matter transforming into radiation or quark-gluon plasma can form a regular black hole. In Ref.~\cite{vertogradov2025epjc}, a method was developed to link the efficiency of matter-to-radiation conversion and the density of emitted energy in the form of electromagnetic waves to the parameters of the initial and new types of matter. We briefly describe this method here.

Consider the gravitational collapse of baryonic matter transitioning into radiation. The total energy density of the matter can be expressed as:
\begin{equation}
\rho = \rho_r + \rho_b,
\end{equation}
where $\rho_r$ represents the radiation energy density and $\rho_b$ corresponds to the baryonic matter density. Importantly, the individual energy-momentum tensors for baryon-like matter and radiation are not conserved separately, but the total energy-momentum tensor must be conserved. This conservation law leads to the following system of equations derived from $T^{ik}_{;k} = 0$:
\begin{eqnarray} \label{eq:system}
\rho_r' r + 2P + 2\rho &=& a(v,r)\rho_r, \nonumber \\
\rho_b' r + 2P + 2\rho &=& -a(v,r)\rho_r,
\end{eqnarray}
where the function $a(v,r)$ determines the efficiency of matter-to-radiation conversion. It can be shown that the efficiency $a$ depends directly on the properties of the new type of matter, specifically its energy density $\rho_n$ and pressure $P_n$, as follows:
\begin{equation} \label{eq:effe}
a(v,r) = \frac{\frac{2}{3}\alpha \rho_n - \left(\frac{8}{3} + 2\alpha\right)P_n - P_n' r}{\alpha \rho_n - P_n}.
\end{equation}
The energy density of the emitted radiation can be described by:
\begin{equation} \label{eq:denrad}
\rho_r = \frac{\alpha \rho_n - P_n}{2\alpha - \frac{2}{3}}.
\end{equation}
Substituting the energy density $\rho_n$ and pressure $P_n$ of the new type of matter into these equations, we obtain the transition from baryon-like matter to the matter described by Eq.~\eqref{eq:density_gen} during gravitational collapse. The efficiency of matter-to-radiation conversion $a$ is given by Eq.~\eqref{eq:effe}, and the radiation energy density is given by Eq.~\eqref{eq:denrad}.

The radiation density $\rho_r$ resulting from this process can be explicitly written as:
\begin{eqnarray} \label{eq:radiation}
\rho_r = \xi \frac{M_0 \beta^{2\alpha}}{(r+\beta)^{2\alpha+3}}, \quad \xi > 0, \nonumber \\
\xi \equiv -\frac{\alpha (\alpha+1)(1-4\alpha^2)}{2\alpha - \frac{2}{3}}.
\end{eqnarray}

Thus, the collapse of matter proceeds through the following stages:

\begin{enumerate}

\item Initially, we consider a star with a baryon-like equation of state $P = \alpha \rho$. At the onset of collapse, the mass function of the star takes the form of the Kiselev mass function:
\begin{equation}
M(v,r) = M_0 + D(v)^{1-2\alpha}.
\end{equation}

\item As the star contracts to a radius $r = r_1$, the baryonic matter begins to transition into a new type of matter, releasing energy in the form of radiation given by Eq.~\eqref{eq:denrad}. At this stage, the mass function of the collapsing cloud becomes:
\begin{equation}
M(v,r) = M_0 + D(v)r^{1-2\alpha} - \alpha(1+2\alpha)\frac{M_0}{\beta^{1-2\alpha}(r+\beta)^{2\alpha-1}}\left[1 + \frac{(1-2\alpha)\beta}{\alpha(r+\beta)} - \frac{(1-2\alpha)\beta^2}{(1+2\alpha)(r+\beta)^2}\right].
\end{equation}
This mass function corresponds to the region within the star where the transition from baryon-like matter to the new type of matter occurs, accompanied by radiation emission described by Eq.~\eqref{eq:denrad}. This process takes place in the spacetime region $0 < r_2 \leq r \leq r_1$. During this phase, the function $D(v)$ decreases with time ($\dot{D} < 0$). Eventually, when the baryonic matter is fully converted ($D(v) \equiv 0$), the collapse enters its third stage.

\item When the matter contracts to a radius $r = r_2$, all matter transitions into the new type described by Eq.~\eqref{eq:denrad}. The mass function corresponding to this region of the collapsing matter is given by:
\begin{equation}
M(v,r) = M_0 - \alpha(1+2\alpha)\frac{M_0}{\beta^{1-2\alpha}(r+\beta)^{2\alpha-1}}\left[1 + \frac{(1-2\alpha)\beta}{\alpha(r+\beta)} - \frac{(1-2\alpha)\beta^2}{(1+2\alpha)(r+\beta)^2}\right].
\end{equation}

\end{enumerate}

In summary, the interior of the collapsing cloud is divided into three regions: an outer region described by baryon-like matter, an intermediate region where baryon-like matter transitions into a new type of matter with accompanying radiation emission, and an inner region where the new type of matter fully forms in the absence of baryonic matter. This final state constitutes the de Sitter core, which prevents the formation of a singularity.

\section{Conclusions}

In this work, we have developed a unified and physically motivated framework for constructing exact dynamical regular black hole solutions arising from the gravitational collapse of baryonic matter. Our approach is based on a systematic regularization of Vaidya-type spacetimes sourced by an anisotropic fluid obeying a generalized polytropic equation of state. The central idea is that, while ordinary barotropic matter inevitably leads to singular geometries, effective nonlinear corrections to the equation of state at high densities can naturally regularize the spacetime and generate a de~Sitter-like core.

Starting from the Kiselev class of solutions, we have introduced a regularization of the energy density profile characterized by a parameter $\beta$, which controls the scale at which deviations from the classical behavior become important. Requiring regularity at the center, finiteness of curvature invariants, and coordinate independence of the equation of state leads to a universal constraint relating $\beta$ to the mass function. As a consequence, the central energy density and the Kretschmann scalar acquire universal finite values that do not depend on the black hole mass, providing a natural resolution of the singularity problem within classical general relativity coupled to effective matter.

A key outcome of our analysis is the emergence of a generalized polytropic equation of state of the form
\[
P = \alpha \rho - \zeta \rho^\gamma ,
\]
which governs the dynamics of the collapsing matter in the regular regime. This equation of state interpolates between ordinary barotropic matter at low densities and an effective vacuum-like behavior at high densities, thereby supporting a de~Sitter core. Importantly, the requirement that the coefficients of this equation of state be independent of spacetime coordinates severely restricts the allowed time dependence of the model parameters, leading to a consistent and predictive framework.

Within this general setting, we have demonstrated that several well-known regular black hole spacetimes arise as particular cases. The Hayward solution appears uniquely in the case of gravitational collapse of a stiff fluid, highlighting the distinguished role of this equation of state. For other types of baryonic matter, the resulting geometries take a more complicated but still fully regular form, which we referred to as generalized Hayward spacetimes. Similarly, by adopting alternative regularization profiles, we showed how Bardeen-like solutions can be obtained, again as special cases of the same underlying polytropic structure. This unifies a broad class of regular black hole models within a single dynamical and physically transparent framework.

Our results suggest that regular black holes need not rely on fundamentally exotic matter sources. Instead, they may be understood as effective descriptions of baryonic matter undergoing gravitational collapse and experiencing nonlinear collective effects at extreme densities. From this perspective, regular black holes represent a plausible endpoint of stellar collapse rather than purely mathematical curiosities. Moreover, the dynamical nature of the solutions presented here makes them suitable for applications in time-dependent scenarios, including black hole formation and evaporation~\cite{Vertogradov:2024fbd}.

Several directions for future work naturally follow from our analysis. It would be of interest to study the stability of the obtained solutions under perturbations, as well as their observational signatures in gravitational-wave emission and black hole shadows. Another important avenue is the investigation of possible microphysical origins of the generalized polytropic equation of state employed here, for instance in the context of dense nuclear matter or effective field theory descriptions of strong interactions. We hope that the framework developed in this paper will serve as a useful starting point for further studies of nonsingular black holes and the physics of gravitational collapse beyond classical singularities.

\section*{Acknowledgments}
Yi Ling is very grateful to Zhangping Yu for helpful discussions and this work is supported in part by the Natural Science Foundation of China (Grant No.~12275275). 

\bibliographystyle{apsrev4-1}
\bibliography{ref}

\end{document}